\documentstyle[11pt,twoside,sympo2011,epsfig,amssymb]{article}
\begin{document}
\title{Messages from the Core: Excited Gravity Modes in the Nearby Blue Supergiant Star Rigel ($\beta$ Ori)}
\author{E. Moravveji$^1$, E. F. Guinan$^2$, A. Moya$^3$, M. Williamson$^4$, F. Fekel$^4$}
\affil{$^1$Institute for Advanced Studies in Basic Sciences, Zanjan, Iran [moravveji@iasbs.ac.ir] \\
$^2$Department of Astronomy, Villanova University, Villanova PA, USA \\ 
$^3$Departamento de Astrofísica, LAEX-CAB (INTA-CSIC), Madrid, Spain \\
$^4$Center of Excellence in Information Systems, Tennessee State University, Nashville, USA} 
\begin{abstract}
Rigel is a bright nearby B8 Ia supergiant star. We observed Rigel for 27.7 days with the MOST satellite and monitored it's 
optical spectrum for over 6 years. Radial velocity (RV hereafter) curve manifestly calls for tens of modes for prewhitening.
We conservatively report on the first 19 significant modes to avoid misdetection of aliases.
The variability periods range from about 75 days down to a day. All radial modes are stable.
We speculate the presence of gravity-dominated mixed-modes excited by $\epsilon-$mechanism from the Hydrogen burning shell 
on top of He burning core. 
\end{abstract}
\section{Introduction}\label{s:intro}
Blue supergiant phase is a very transient, short-lived evolutionary phase of massive stars ($M \gtrsim 15 M_\odot$), and hence
the study of the internal structure of these core-collapse supernova progenitors could bring important constraints on 
the mass, size, composition and rotation rate of the He core prior to the collapse.
This, however, depends critically on how different mixing mechanisms such as overshooting, semi-convection and rotation are treated 
during the main sequence (MS) phase of massive stars (Heger et al. 2000; Noels et al. 2010).
In this respect, we find Rigel an excellent testbed for such investigations.

Saio et al. (2006) detected 48 pulsation modes in the MOST post-MS target HD 163899 (B2 Ib/II). They explained the excitation 
of observed g-modes in terms of mode reflection from the deep lying intermediate convective zone (ICZ) when they arrive at the appropriate phase to 
the edge of the ICZ. Godart et al. (2009) theoretically investigated the destructive effects of mass loss and core overshooting
on the location and extent of the ICZ, and the asteroseismic potential of this class of stars. Since a limited number of modes can 
survive the huge radiative damping in the core, they propose to artificially impose the inner boundary condition at the base of ICZ. 
According to Saio (2011) expected instabilities for stars like Rigel are radial modes in addition to $l=1$ and 2 convection g$^-$ modes. 
However, Gautschy (2009) predicts no instability for stars with $3.95 \lesssim \log T_{\mbox{\scriptsize eff}} \lesssim 4.15$ 
(Rigel lies in this gap).
This is why asteroseismology of Rigel could be model-dependent and demanding.

Several observed quantities of Rigel which impose valuable constraints on the equilibrium model of the star are already measured. 
They are the Hipparcos parallax $\pi=3.79\pm0.34$ mas, effective temperature $T_{\mbox{\scriptsize eff}}=12100\pm150$ K, surface gravity 
$\log g=1.75\pm0.10$, luminosity $\log (L/L_\odot)\simeq5.07\pm0.10$, metalicity [M/H]=$-0.06\pm0.10$ (Przybilla et al. 2006), 
surface He abundance $Y_{\mbox{\scriptsize s}}=0.32\pm0.04$, $v\sin i\approx25\pm3$ km s$^{-1}$ (Przybilla et al. 2010; Simon-Diaz et al. 2010), and
limb darkened angular diameter $\theta_{\mbox{\scriptsize LD}}=2.75\pm0.01$ mas. 
Moreover, Rigel's variability in the H$\alpha$ line profile (Kaufer et al. 1996; Morisson et al. 2008) and other metal-line equivalent widths due to 
non-radial pulsations (Kaufer et al. 1997) and mass loss ($\dot{M}=1.5\pm0.4\times10^{-7} M_\odot$ yr$^{-1}$, Chesneau et al. 2010) are already published.
These observed values are later tried to be reproduced when modeling the star.
When modeling the equilibrium state of Rigel (section \ref{s:graco}) we attempted to match the above mentioned observed quantities with the modeled 
quantities.
\section{Variability Time Scales From Radial Velocity Monitoring}
Photometry of Rigel was carried out with MOST (Walker et al. 2003) satellite for 27.7 days from 15 November to 13 December 2009. 
Since our photometry time-baseline is of the same order of the long-period modes no definite frequencies are found in Lomb-Scargle periodogram.
The RV monitoring, in return, extends over 6 years but suffers from daily and annual gaps. 
We detected 19 significant harmonics using \texttt{SigSpec} software (Reegen 2007) above sig$=$15 threshold; 
the results are in complete agreement with those of \texttt{Perod04} software (Lenz \& Breger 2005) with SNR$\gtrsim$4.6.
The extensive list of harmonics will be presented in Moravveji et al. (in preparation).
The two lowest frequencies occur at 0.0152 and 0.0134 d$^{-1}$ with semi-amplitudes of 1.126 and 0.838 km s$^{-1}$, respectively, and
the highest frequency occurs at 0.8203 d$^{-1}$. 
\section{Equilibrium Modeling and Non-adiabatic Radial and Non-radial Pulsations }\label{s:graco} 
Stellar structure and evolution modeling which includes rotational mixing (Heger et al. 2000), mass loss, and element diffusion subject to
constraints listed in section \ref{s:intro} is carried out with the \texttt{MESA} code (Paxton et al. 2011). 
Schwarzschild criterion for convection with $\alpha_{\mbox{\scriptsize MLT}}=1.6$ is used. 
So, the main source of energy is a slight He burning in the core followed by H burning in shell. 
But models with Ledoux convective criterion have no He burning core (Godart et al. 2009). 

Since non-adiabaticity can play a critical role in excitation and damping of modes in massive stars, we treat modal stability analysis employing
Granada Oscillatin Code (\texttt{GraCo}, Moya et al. 2004). 
For the input equilibrium model from \texttt{MESA}, first we scan for $0\leq\ell\leq3$ modes in adiabatic frame, 
and next we solve the full non-adiabatic equations.
Even without imposing the inner boundary condition at the base of the ICZ (suggested by Godart et al. 2009), there are several unstable high order 
low-degree \textit{mixed} gravity modes as eigensolutions. 
Fundamental radial mode and its overtones are all stable.
The radiative layer below the ICZ where H-burning takes place shows significant contribution in eigenfunction of excited modes.
After analyzing the work integral $W$ and its derivative $dW/d\ln T$, we speculate that the pulsation driving arises from the 
$\epsilon-$mechanism (Unno et al. 1989).
This is powered by the sensitivity of the CNO burning network to the temperature perturbations below the ICZ of $\log T\approx7.7$
which dominates over $\kappa-$mechanism around
Iron bump of opacity located at $\log T\approx5.2$.
If securely proven, this approach will open a venue towards studies of near-core condition in progenitors of SN II progenitors. 
This part of the study is still under progress.
\acknowledgments{Ehsan Moravveji appreciates the grant he received to attend the conference. This work is supported by NASA/MOST grant NNX09AH28G.}


\end{document}